\documentclass[%
reprint,
superscriptaddress,
 amsmath,amssymb,
prl,
]{revtex4-1}
\usepackage{mathrsfs}
\usepackage[pdftex]{graphicx}
\usepackage{dcolumn}
\usepackage{bm}
\usepackage{longtable}
\usepackage{color,hyperref}
\usepackage{tikz}
\usepackage{longtable}
\hypersetup{colorlinks,breaklinks,linkcolor=blue,urlcolor=blue,anchorcolor=blue,citecolor=blue}
\usepackage{comment}
\usepackage{multirow}            
\usepackage{physics}            

\begin{document}

\preprint{APS/123-QED}

\title{Evidence for rigid triaxial deformation in $^{76}$Ge from a model-independent analysis}

\author{A. D. Ayangeakaa}
 \email[Corresponding authors: ]{ayangeak@usna.edu,\\ rvfj@email.unc.edu}
 \affiliation{Department of Physics, United States Naval Academy, Annapolis, Maryland 21402, USA}

 \author{R. V. F. Janssens}
\email[Corresponding authors: ]{ayangeak@usna.edu,\\ rvfj@email.unc.edu}

\affiliation{Department of Physics and Astronomy, University of North Carolina at Chapel Hill, Chapel Hill, North Carolina 27599, USA}
\affiliation{Triangle Universities Nuclear Laboratory, Duke University, Durham, North Carolina 27708, USA}

\author{S. Zhu}
\altaffiliation[Present Address: ]{National Nuclear Data Center, Brookhaven National Laboratory, Upton, New York 11973-5000, USA}
\affiliation{Physics Division, Argonne National Laboratory, Argonne, Illinois 60439, USA}

\author{D. Little}
\affiliation{Department of Physics and Astronomy, University of North Carolina at Chapel Hill, Chapel Hill, North Carolina 27599, USA}
\affiliation{Triangle Universities Nuclear Laboratory, Duke University, Durham, North Carolina 27708, USA}
\author{J. Henderson}
\affiliation{Lawrence Livermore National Laboratory, Livermore, California 94550, USA}

\author{C. Y. Wu} 
\affiliation{Lawrence Livermore National Laboratory, Livermore, California 94550, USA}

\author{D. J. Hartley}
 \affiliation{Department of Physics, United States Naval Academy, Annapolis, Maryland 21402, USA}
 
 \author{M. Albers}
\affiliation{Physics Division, Argonne National Laboratory, Argonne, Illinois 60439, USA}

\author{K. Auranen}
\affiliation{Physics Division, Argonne National Laboratory, Argonne, Illinois 60439, USA}

\author{B. Bucher} 
 \altaffiliation[Present Address: ]{Idaho National Laboratory, Idaho Falls, Idaho 83415, USA}
\affiliation{Lawrence Livermore National Laboratory, Livermore, California 94550, USA}
\author{M. P. Carpenter}
\affiliation{Physics Division, Argonne National Laboratory, Argonne, Illinois 60439, USA}

\author{P. Chowdhury}
\affiliation{Department of Physics and Applied Physics, University of Massachusetts Lowell, Lowell, Massachusetts 01854, USA}
\author{D. Cline}
\affiliation{Department of Physics and Astronomy, University of Rochester, Rochester, New York 14627, USA}

\author{H. L. Crawford}
\affiliation{Nuclear Science Division, Lawrence Berkeley National Laboratory, Berkeley, California 94720, USA}

\author{P. Fallon}
\affiliation{Nuclear Science Division, Lawrence Berkeley National Laboratory, Berkeley, California 94720, USA}
\author{A. M. Forney}
\affiliation{Department of Chemistry and Biochemistry, University of Maryland, College Park, Maryland 20742, USA}

\author{A. Gade}
\affiliation{National Superconducting Cyclotron Laboratory, Michigan State University, East Lansing, Michigan 48824, USA}
\affiliation{Department of Physics and Astronomy, Michigan State University, East Lansing, Michigan 48824, USA}

\author{A. B. Hayes}
\affiliation{Department of Physics and Astronomy, University of Rochester, Rochester, New York 14627, USA}

\author{F. G. Kondev}
\affiliation{Physics Division, Argonne National Laboratory, Argonne, Illinois 60439, USA}

\author{Krishichayan}
\affiliation{Triangle Universities Nuclear Laboratory, Duke University, Durham, North Carolina 27708, USA}
\affiliation{Department of Physics, Duke University,Durham, North Carolina 27708, USA}

\author{T. Lauritsen}
\affiliation{Physics Division, Argonne National Laboratory, Argonne, Illinois 60439, USA}

\author{J. Li}
\affiliation{Physics Division, Argonne National Laboratory, Argonne, Illinois 60439, USA}

\author{A. O. Macchiavelli}
\affiliation{Nuclear Science Division, Lawrence Berkeley National Laboratory, Berkeley, California 94720, USA}

\author{D. Rhodes}
\affiliation{National Superconducting Cyclotron Laboratory, Michigan State University, East Lansing, Michigan 48824, USA}
\affiliation{Department of Physics and Astronomy, Michigan State University, East Lansing, Michigan 48824, USA}

\author{D. Seweryniak}
\affiliation{Physics Division, Argonne National Laboratory, Argonne, Illinois 60439, USA}
\author{S. M. Stolze}
\affiliation{Physics Division, Argonne National Laboratory, Argonne, Illinois 60439, USA}

\author{W. B. Walters}
\affiliation{Department of Chemistry and Biochemistry, University of Maryland, College Park, Maryland 20742, USA}

\author{J. Wu}
\affiliation{Physics Division, Argonne National Laboratory, Argonne, Illinois 60439, USA}

\date{\today}

\begin{abstract} 
An extensive, model-independent analysis of the nature of triaxial deformation in $^{76}$Ge, a candidate for neutrinoless double-beta ($0\nu\beta\beta$) decay, was carried out following  multi-step Coulomb excitation. Shape parameters deduced on the basis of a rotational-invariant sum-rule analysis provided considerable insight into the underlying collectivity of the ground-state and $\gamma$ bands. Both sequences were determined to be characterized by the same $\beta$ and $\gamma$ deformation parameter values. In addition, compelling evidence for low-spin, rigid triaxial deformation in $^{76}$Ge was obtained for the first time from the analysis of the statistical fluctuations of the quadrupole asymmetry deduced from the measured $E2$ matrix elements. These newly determined shape parameters are important input and constraints for calculations aimed at providing, with suitable accuracy, the nuclear matrix elements relevant to $0\nu\beta\beta$.

\end{abstract}
\maketitle

Neutrinoless double-beta ($0\nu\beta\beta$) decay is one  of  the most promising experimental  techniques  capable  of  probing  the  fundamental properties  of  the  neutrino~\cite{Gomez-Cadenas}. The observation of this rare weak-interaction process would signal a violation of total lepton number conservation and establish the Majorana nature of the neutrino; e.g., that the neutrino is its own antiparticle. In addition, the measured $0\nu\beta\beta$ half-life would potentially provide experimental access to the absolute neutrino mass scale, provided that the nuclear matrix elements (NME) mediating the decay are reliably known. However, results of nuclear structure calculations of the NMEs differ by up to a factor of three~\cite{Engel_2017,PhysRevLett.109.042501}, depending on the methodology. This translates into an order of magnitude variation in the decay lifetime. Experimental input from a nuclear structure perspective to constrain these calculations is, thus, essential as this would allow models to be selected or developed based on reproducible benchmarking criteria.

In this regard, wavefunctions of leading $0\nu\beta\beta$ candidates have been probed in a campaign of experiments utilizing transfer reactions to determine nucleon occupancies in a consistent way~\cite{Schiffer-PRL.100.112501,Kay-PRC.79.021301}. These studies have provided critical information for comparison with theory, especially on contributions to the wavefunctions from competing single-particle orbitals. In much the same way, recent inelastic neutron scattering measurements have provided spectroscopic information on the structure of low-lying states~\cite{PhysRevC.95.014327,PhysRevC.99.014313}, with implications on the kinematic phase space available for the $0\nu\beta\beta$ process. However, all these studies lack the required level of sensitivity to collective degrees of freedom which have been shown to significantly impact the calculated NMEs. For example, deformation due to quadrupole correlations has been found to reduce the calculated NME strengths~\cite{Menendez:2008jf,PhysRevC.83.034320,PhysRevC.87.064302,PhysRevC.94.014306}, especially when the parent and daughter nuclei have different shapes. On the other hand, NMEs are enhanced when the deformations are similar~\cite{Tomas-PRL.105.252503,PhysRevC.94.014306}. Moreover, it has been demonstrated  that the calculated NMEs are largest when spherical symmetry is assumed in both parent and daughter nuclei~\cite{PhysRevC.91.024316}. Hence, the proper treatment of deformation and the role of axial asymmetry are essential for reliably calculating the NMEs for $0\nu\beta\beta$ decay.

Experimental investigations aiming to observe $0\nu\beta\beta$ decay are underway. Among the isotopes considered, the $0\nu\beta\beta$ decay of $^{76}$Ge into $^{76}$Se possesses high discovery potential and is currently the focus of the \textsc{Gerda}~\cite{Collaboration2017,PhysRevLett.111.122503} and \textsc{Majorana Demonstrator} ~\cite{PhysRevLett.120.132502,majorana-coll} collaborations. For this parent-daughter pair, theoretical NME predictions differ by factors of $\sim$$2$-$3$ between different shell-model approaches and by as much as $\sim$$5$-$6$ between quasiparticle random phase approximation (QRPA) and energy density functional (EDF) ones~\cite{PhysRevC.83.034320}.

From a nuclear structure point of view, the low-lying structure of $^{76}$Ge is a subject of significant interest, since it has been suggested to represent a rare example of rigid triaxiality at low spin, including the ground state~\cite{Toh2013}. For decades, the experimental observation of such rigid triaxiality has remained a major challenge. Nuclear triaxiality, which is expressed in terms of the asymmetry parameter $\gamma$, has traditionally been described using two major phenomenological models. The $\gamma$-rigid triaxial rotor model of Davydov and Filippov (DF)~\cite{DAV} assumes a collective potential with a stable minimum at a finite value of $\gamma$. In contrast, the $\gamma$-soft rotor model of Wilets and Jean (WJ)~\cite{Wilets-Jean1956} incorporates a $\gamma$-independent collective potential. 

As mentioned above, experimental evidence for low-spin rigid-triaxial deformation was recently proposed, based on the energy pattern of the low-spin structure of $^{76}$Ge~\cite{Toh2013}, where the phase of the odd-even staggering in the  $\gamma$ band is consistent with DF model predictions. Since then, a similar 
pattern was observed in $^{78}$Ge as well~\cite{PhysRevLett.120.212501}. 
Note that the phase of the energy staggering in the  DF model is opposite to that of the WJ one, and has been suggested to be a measure of the degree of stiffness or softness of the $\gamma$ deformation~\cite{PhysRevC.76.024306}. It is also important to note that, while the phase of the staggering in $^{76}$Ge is consistent, the amplitude is less than the DF model prediction for $\gamma = 30^\circ$.  This discrepancy has resulted in a range of theoretical investigations with varying, and sometimes conflicting, conclusions. For instance, while calculations performed within the relativistic Hartree--Bogoliubov (RHB) approach with a universal functional~\cite{PhysRevC.78.034318} predict dynamic ($\gamma$-soft) deformation for $^{76}$Ge~\cite{PhysRevC.89.044325}, microscopic calculations using the multi-quasiparticle triaxial projected shell model (TPSM)~\cite{PhysRevC.89.014328} and the symmetry-conserving configuration mixing methods based on the Gogny D1S interaction~\cite{Rodr_guez_2017} require a fixed (rigid) triaxial deformation of $\gamma \approx 30^\circ$ to reproduce the structure of $^{76}$Ge. Similarly, theoretical investigations with the proton-neutron variant of the interacting boson model (IBM2)~\cite{Zhang_2013} are able to reproduce the energy staggering of the $\gamma$ band. On the other hand, phenomenological pairing-plus-quadrupole shell-model calculations~\cite{PhysRevC.78.044320} account for the level structure of $^{76}$Ge without invoking $\gamma$ deformation.

In this letter, a model-independent study of the quadrupole triaxial degree of freedom, based on measured $E2$ transition matrix elements in $^{76}$Ge, is presented. The $E2$ properties of nuclear states, determined via Coulomb excitation, are the most sensitive measure of quadrupole collectivity and provide a more direct indication of triaxiality than level energies or branching ratios used earlier to investigate the nature of $\gamma$ deformation in this nucleus. It is noted that this process was used to investigate $^{76}$Ge in the earlier work of Toh {\it et al.}~\cite{Toh2001}, but with limited population of the relevant states. The unique and complete set of $E2$ matrix elements obtained in the present study now permits a model-independent characterization of the low-spin structure of $^{76}$Ge. Model independency is obtained from the direct application of the rotational invariant sum-rule method~\cite{KK1,KK2}. The latter is model independent within the general framework of the collective model; it enables the determination of the deformation parameters in the intrinsic frame from the measured $E2$ matrix elements without any assumption about the nuclear shape. This approach was recently followed in the case of $^{76}$Se, the $0\nu\beta\beta$-decay daughter of $^{76}$Ge, to characterize the degree of triaxiality of its ground state~\cite{PhysRevC.99.054313}. The comprehensive data set obtained in the present work allows the expansion of this sum-rule analysis further, showing not only that the $^{76}$Ge ground-state and $\gamma$ bands exhibit the same triaxial deformation, but that, in addition, rigid triaxiality with an asymmetry  close to $30^\circ$  is derived for the three lowest states in the nucleus.

\begin{figure}[b!]
\hspace{-1.em}
\includegraphics[width=\columnwidth]{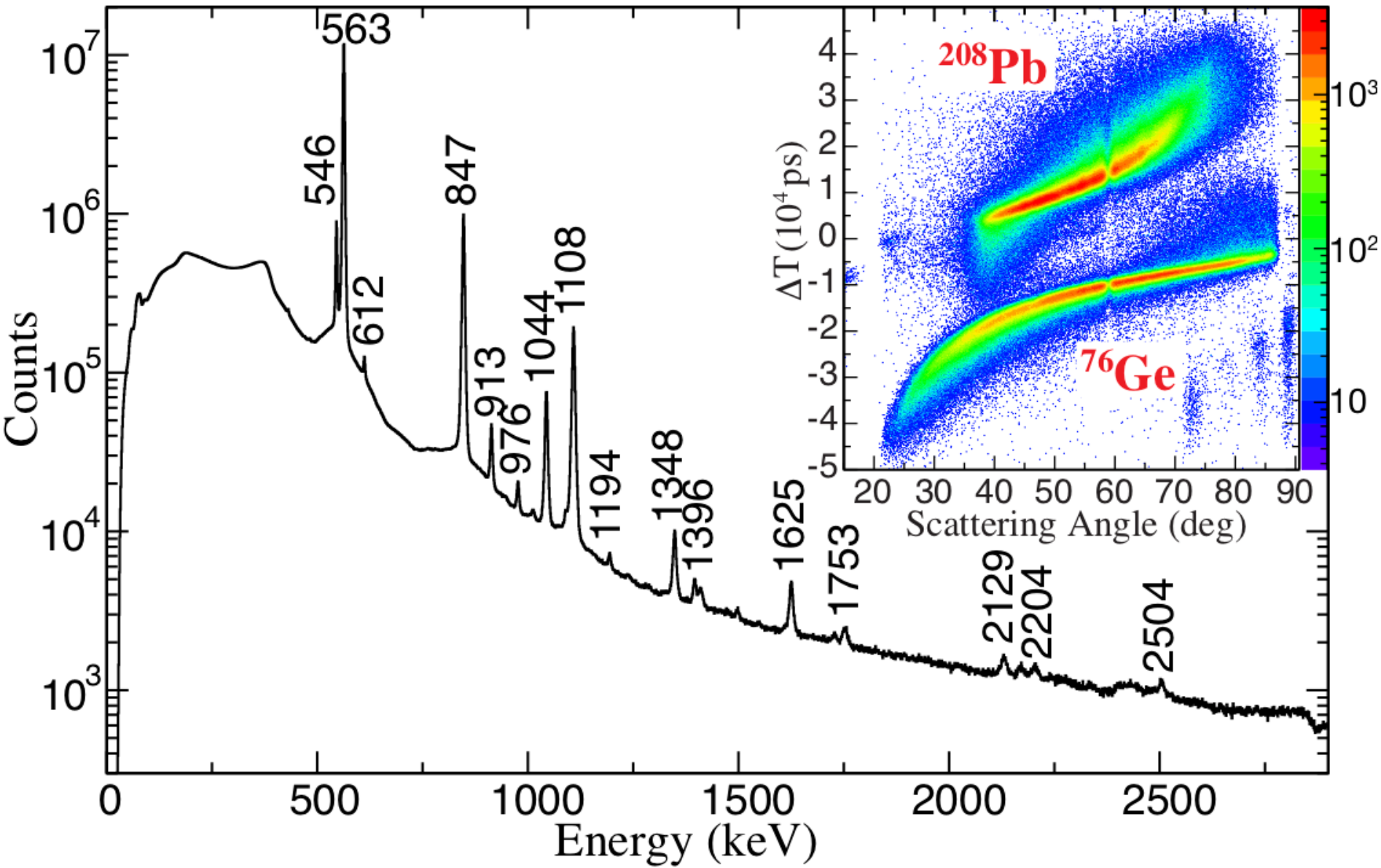}
\caption{\label{fig:specall} (Color online) Untracked Doppler-corrected $\gamma$-ray spectrum obtained in kinematic coincidence with $^{76}$Ge ions. The insert illustrates the performance of the CHICO2 array in discriminating between the projectile and target nuclei.} 
\end{figure}

\begin{figure}[b!]
\includegraphics[width=1.0\columnwidth]{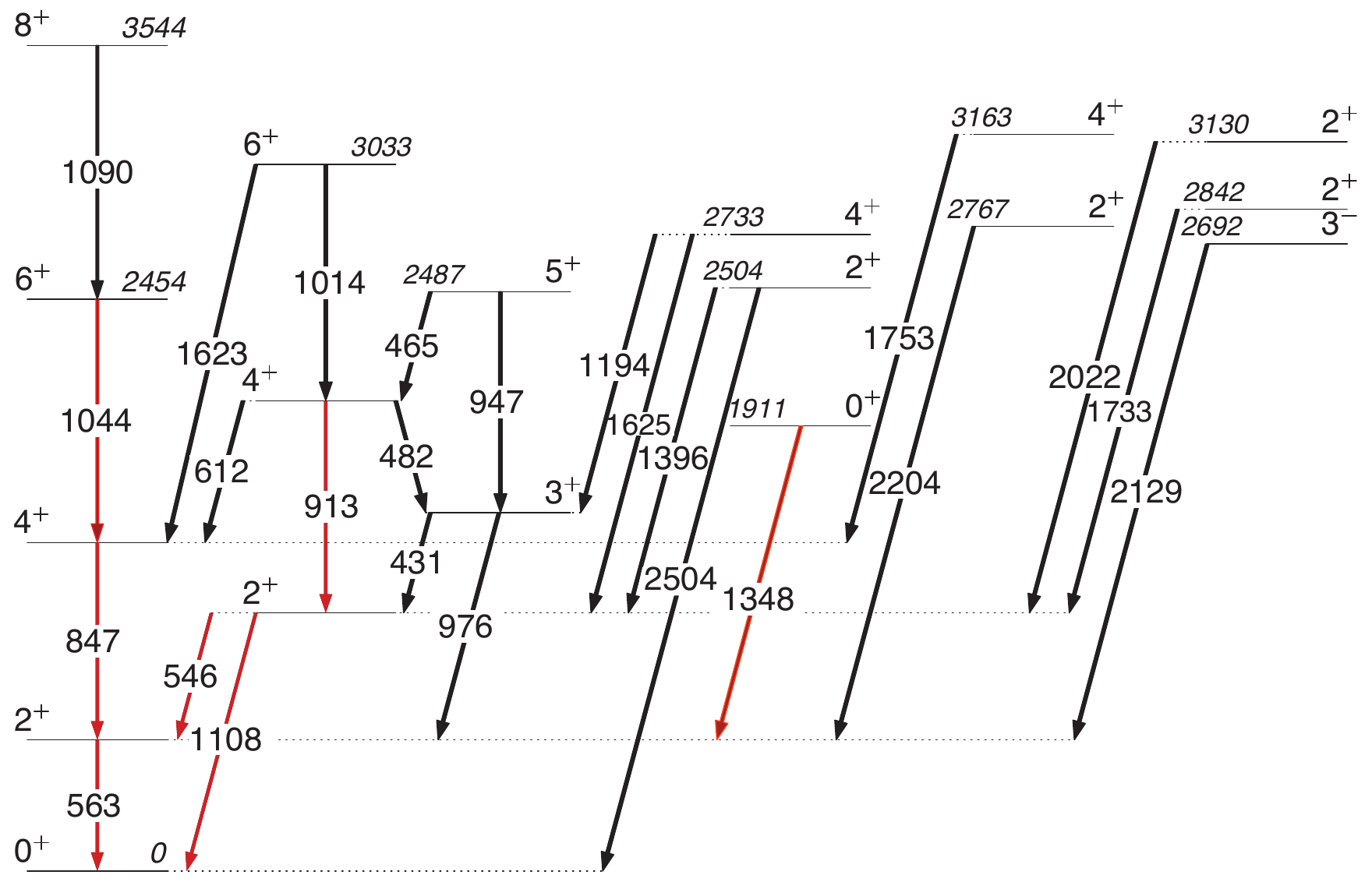}
\caption{\label{fig:levelsch} (Color online) A partial level scheme with all the levels populated in the present Coulomb excitation measurement. Transitions in red are those observed in the previous measurement of this type~\cite{Toh2001}.}
\end{figure}

The present measurements were performed using the same experimental approach as that described in Ref.~\cite{Ayangeakaa2016} for Coulomb excitation of $^{72}$Ge. Hence, the experimental procedure and  analysis methods are only briefly summarized here. Two separate experiments with three beam energies were performed at Argonne National Laboratory. In both experiments, $^{76}$Ge ion beams from the ATLAS accelerator bombarded a 0.5 mg/cm$^2$-thick $^{208}$Pb target, sandwiched between a 6 $\mu \mathrm{g}/\mathrm{cm}^2$  Al front layer and a 40  $\mu \mathrm{g}/\mathrm{cm}^2$ C backing. The de-excitation $\gamma$ rays were detected by the tracking array, GRETINA~\cite{Paschalis201344} in kinematic coincidence with scattered reaction products recorded with the CHICO2 array of position-sensitive parallel plate avalanche counters~\cite{Wu2016}.
The first experiment utilized a sub-barrier beam energy of 304 MeV and 7 GRETINA modules (28 Ge crystals). For the second, two beam energies of 291 and 317 MeV  were employed along with 11 GRETINA modules (42 crystals). A summed $\gamma$-ray spectrum, after Doppler correction, measured in coincidence with the scattered $^{76}$Ge projectiles is presented in Fig.~\ref{fig:specall}. The inset depicts a two-dimensional histogram of differences in the time of flight ($\Delta T_{\mathrm{tof}}$) between reaction partners versus scattering angle, $\theta$, demonstrating the clear separation between projectile and target nuclei. A partial level scheme, incorporating all the $^{76}$Ge states populated in this work, is displayed in Fig.~\ref{fig:levelsch}; the red-colored transitions are those seen in the prior Coulomb excitation measurement~\cite{Kot1990646}. The present level scheme confirms results from earlier works (see Ref.~\cite{Toh2013}, for example). 

\begingroup
\begin{table}
\caption{\label{tab2} $E2$ matrix elements for $^{76}$Ge obtained from the present analysis and comparisons with previous measurements. Note that not all matrix elements corresponding to the levels shown in Fig.~\ref{fig:levelsch} are given here. The complete set will be provided in a forthcoming publication~\cite{ayangeakaa2}.}

\begin{tabular}{crrrrrrrr}
\hline\hline
\multicolumn{1}{c}{\multirow{2}{*}{$I_i^\pi \rightarrow I_f^\pi$}} & \multicolumn{6}{c}{$\bra{I_i} |\mathcal{M}(E2)| \ket{I_f}$ $(e\mathrm{b})$}                                                                                                                                       \\ \cline{3-7} 
\multicolumn{1}{c}{}                     && \multicolumn{1}{c}{This Work}     && \multicolumn{1}{c}{Ref.~\cite{Toh2001}} && \multicolumn{1}{c}{Refs.~\cite{PhysRevC.22.2420,PhysRevC.22.1530}} \\ \hline
$0_1^+\rightarrow 2_1^+$		&&	0.526(2)	&&	0.522(4)		&&	0.550(3) \\
$0_1^+\rightarrow 2_2^+$		&& 0.089(3)	&&	$0.069(10)$	&&	$\bigl|0.081(14)\bigr|$	\\
$0_1^+\rightarrow 2_3^+$		&& $0.061(3)$  			&&			&&\\
$0_1^+\rightarrow 2_4^+$		&& $0.054(4)$  			&&			&&\\
$0_1^+\rightarrow 2_5^+$		&& $0.023(6)$  			&&			&&\\
$2_1^+\rightarrow 2_1^+$		&&   $-0.24(2)$	&&	$-0.14(4)$ &&$-0.19(6)$\\
$2_1^+\rightarrow 2_2^+$		&& $0.535_{-0.007}^{+0.003}$	&&	$0.54(3)$		&&	$\bigl|0.71(7)\bigr|$	\\ 
$2_1^+\rightarrow 2_3^+$		&& $-0.126_{-0.004}^{+0.006}$	&&			&&			     \\
$2_1^+\rightarrow 2_4^+$		&& $0.022_{-0.005}^{+0.008}$	&&			&&			     \\
$2_1^+\rightarrow 2_5^+$		&& $-0.048_{-0.007}^{+0.002}$	&&			&&			     \\
$2_1^+\rightarrow 3_1^+$		&& $0.082(5)$	&&		&&		\\ 
$2_1^+\rightarrow 4_1^+$		&&	$0.795(5)$	&&	0.71(4)		&&	0.77(4)   \\
$2_1^+\rightarrow 4_2^+$		&&   	$-0.22_{-0.03}^{+0.05}$	&&   0.10(2)   		&&	 		\\
$2_2^+\rightarrow 2_2^+$		&&   	$0.26_{-0.05}^{+0.02}$	&&	$0.28(6)$	&&	 			\\
$2_2^+\rightarrow 3_1^+$		&& $0.52_{-0.04}^{+0.02}$	&&		&&				\\
$2_2^+\rightarrow 4_2^+$		&&   	$0.472(6)$	&&	$0.56(2)$		&&	 		\\
$4_1^+\rightarrow 4_1^+$		&&   $-0.26_{-0.07}^{+0.01}$	&&	$-0.01(5)$		&&	 		\\
$4_1^+\rightarrow 6_1^+$		&&	$1.11_{-0.02}^{+0.03}$	&&	0.87(2)		&&	 	       \\
$6_1^+\rightarrow 8_1^+$		&&	$1.25_{-0.10}^{+0.07}$	&&				&&	 		\\           $6_1^+\rightarrow 6_1^+$		&&   $-0.23_{-0.04}^{+0.09}$	&&		&&	        		\\   
$3_1^+\rightarrow 5_1^+$		&&  $0.9_{-0.6}^{+0.4}$	&&		&&						\\  
$4_2^+\rightarrow 3_1^+$		&&  $0.64_{-0.07}^{+0.03}$	&&		&&						\\  
$4_2^+\rightarrow 5_1^+$		&&  $0.9_{-0.2}^{+0.7}$	&&		&&						\\  
$4_2^+\rightarrow 6_2^+$		&&  0.49(3)	&&		&&						\\  
$6_2^+\rightarrow 5_1^+$		&&  $-0.74_{-0.08}^{+0.10}$	&&		&&						\\  
$3_1^+\rightarrow 3_1^+$		&& $0.13_{-0.10}^{+0.08}$  	&&			&&				\\ 
$4_2^+\rightarrow 4_2^+$		&&   	$-0.24^{+0.08}_{-0.04}$		&&			&&	 		\\
$4_1^+\rightarrow 2_2^+$		&&  $0.09(2)$ 			&&	$-0.11(1)$		&&	\\
$4_1^+\rightarrow 3_1^+$		&&  	$-0.44_{-0.05}^{+0.08}$		&&		&&	\\
$4_1^+\rightarrow 4_2^+$		&&  $0.61(1)$	&&	$-0.10(3)$	&&				\\ 
$4_1^+\rightarrow 5_1^+$		&&  	$-0.08_{-0.05}^{+0.09}$		&&		&&	\\
$4_1^+\rightarrow 6_2^+$		&&  $-0.186_{-0.005}^{+0.030}$	&&		&&	                 	\\ 
$6_1^+\rightarrow 4_2^+$		&&  $0.35_{-0.03}^{+0.05}$	&&	0.21(4)	&&						\\
\hline \hline
\end{tabular}
\end{table}
\endgroup

The measured $\gamma$-ray intensities were analyzed using the semi-classical, coupled-channel, Coulomb excitation least-squares search code, \textsc{gosia}~\cite{gosia}. To enhance the sensitivity to the matrix elements and exploit the dependence of the excitation probability on the particle scattering angle, the data from each of the three beam energies were partitioned into seven angular subsets. These were analyzed both independently and combined to check for consistency. In addition, known spectroscopic data  such as lifetimes, branching and $E2/M1$ mixing ratios~\cite{Singh199563,PhysRevC.95.014327} were included as constraints in the multi-dimensional fit of the relevant parameters. The final set of matrix elements, which best reproduces the measured $\gamma$-ray yields and the available literature data, as well as their associated uncertainties is displayed in Table~\ref{tab2}. The absolute values and signs of 103 $E2$ and $M1$ reduced matrix elements were determined with sufficient accuracy for a meaningful determination of the rotational invariants (see below). The present results are in general agreement with those of prior measurements of this kind~\cite{Toh2001,PhysRevC.22.2420,PhysRevC.22.1530}. For the purpose of this discussion, however, only a subset of the relevant $E2$ matrix elements are tabulated.

A model-independent analysis of the deformation of $^{76}$Ge at low spin was carried out using the non-energy-weighted sum rules technique described in Refs.~\cite{KK1,KK2,SREBRNY200625}.  
 In this framework, the expectation values of invariant products of the collective $E2$ operator determine the nuclear charge distribution via an intermediate-state expansion over the experimental $E2$ matrix elements. This allows the collective quadrupole invariants to be expressed as functions of the two charge deformation parameters, $Q$ and $\delta$.  Specifically, the quadrupole invariant $\left < Q^2 \right >$ provides an average measure of the static and dynamic intrinsic quadrupole deformation of a charged ellipsoid; i.e., the overall deviation from sphericity. It is equivalent to the elongation parameter $\beta$ in the Bohr Hamiltonian~\cite{bohr1998nuclear}. Similarly, the quadrupole asymmetry $\left < \mathrm{cos}\; 3\delta \right >$ describes the departure from axial symmetry; the parameter $\expval{\delta}=\tfrac{1}{3}\, \mathrm{arccos}\, (\expval{\cos 3\delta})$ is analogous to the collective-model asymmetry angle $\gamma$. Furthermore, higher-order invariants can similarly be constructed using the $J=0,2,4$ coupling schemes~\cite{gosia,SREBRNY200625}.  In particular, the relative stiffness in the $\left < Q^2 \right >$ and $\left < \mathrm{cos}\; 3\delta \right >$ invariants can be determined by evaluating their statistical fluctuations, or dispersion, $\sigma \left < Q^2 \right >$ and $\sigma \left < \mathrm{cos}\; 3\delta \right >$ over a range of reduced matrix elements. The latter quantities allow for the unambiguous determination of whether a nucleus is rigidly deformed and/or rigidly asymmetric.

\begin{figure}[t!]
\begin{center}
\includegraphics[width=\columnwidth]{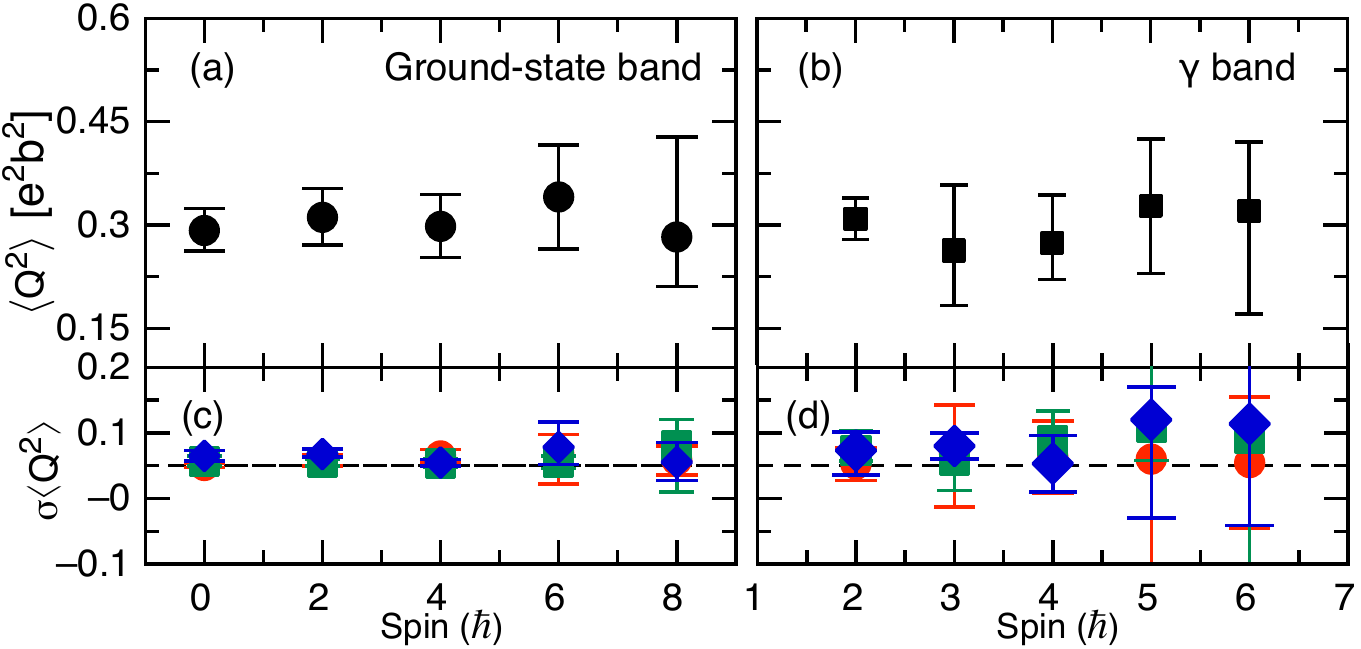}
\caption{(Color online) Magnitude of the quadrupole invariant, $\left < Q^2 \right >$,  for the ground-state  and $\gamma$ bands. The bottom figures present the statistical fluctuation of the quadrupole deformation, $\sigma \left < Q^2 \right >$. Here, red (circle), blue (diamond) and green (square) correspond to $J=0,2,4$ couplings, respectively~\cite{gosia,SREBRNY200625} (see text for details).} 
\label{FIG3lab}
\end{center}
\end{figure}

\begin{figure}[b!]
\begin{center}
\includegraphics[width=\columnwidth]{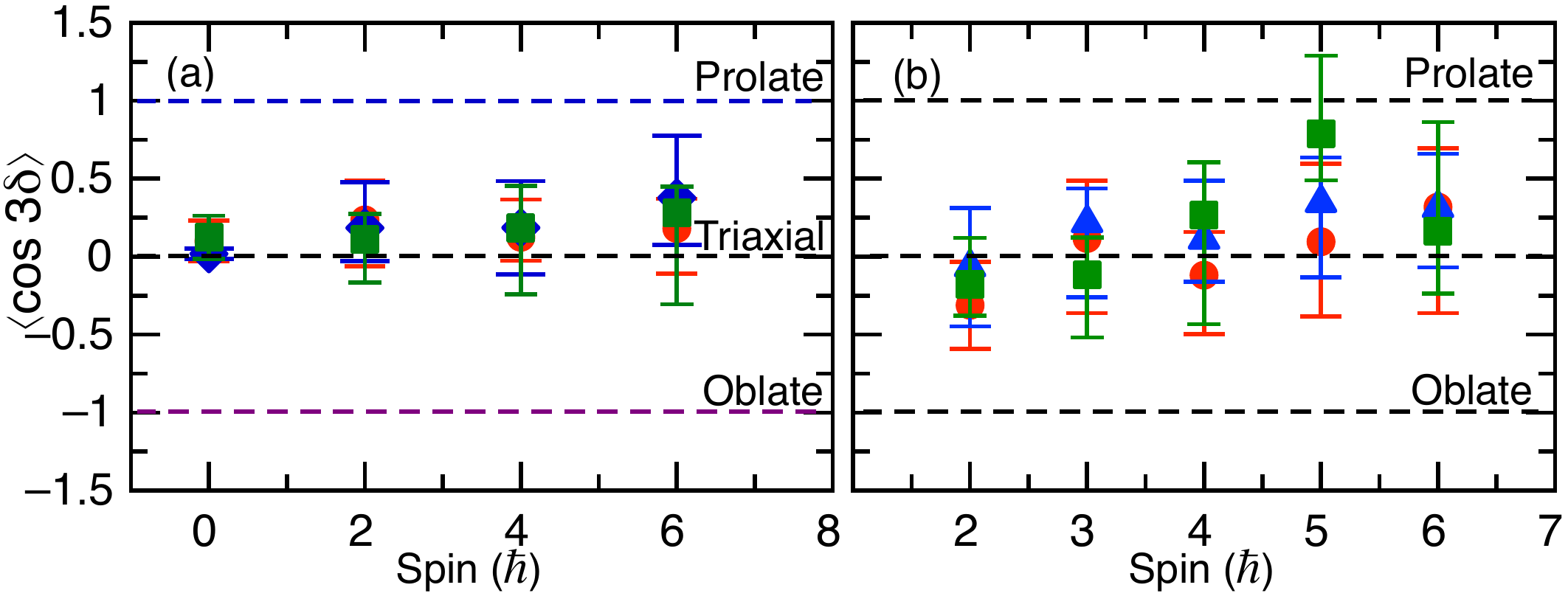}
\caption{(Color online) Expectation values of the quadrupole asymmetry $\left < \mathrm{cos}\; 3\delta \right > $ for (a) the ground-state band and (b) the $\gamma$ band. The same color convention as in FIG.~\ref{FIG3lab} has been used.}
\label{FIG4lab}
\end{center}
\end{figure}
Figures~\ref{FIG3lab}(a) and (b) present the expectation values of $\left < Q^2 \right >$ for levels within the ground-state and $\gamma$ bands, derived from the measured $E2$ matrix elements. From Fig.~\ref{FIG3lab}, it is clear that the $\expval{Q^2}$ values are the same in both sequences and that they are also constant with spin within errors. This spin independence of the $\left < Q^2 \right >$ invariant implies the presence of strong correlations between the low-lying states, as expected in view of the rotational-like behavior exhibited by the $E2$ matrix elements. Furthermore, the notable similarity of $\left < Q^2 \right >$ values for the ground-state and $\gamma$ bands indicates a uniform deformation over the entire spin range that averages a value of $\sim 0.30$ e$^2$b$^2$, corresponding to an average quadrupole deformation of $\beta \approx 0.28$. Here, the transformation $\beta=4\pi\sqrt{\expval{Q^2}}/3ZR^2$ has been applied with  $R=1.2A^{1/3}$, $Z$ and $A$ being the atomic and mass numbers~\cite{SREBRNY200625}. This value agrees with recent symmetry conserving configuration mixing (SCCM) calculations with the Gogny D1S interaction that predict $\beta =0.3$~\cite{Rodr_guez_2017}. Moreover, the overall constancy of the $\expval{Q^2}$ values over the entire spin range in both sequences affirms the general conclusion that the ground-state and $\gamma$ bands are 
built on the same deformation.  The dispersion 
of the quadrupole deformation, $\sigma \left < Q^2 \right >$, for the ground-state and $\gamma$ bands can be found in Figs.~\ref{FIG3lab}(c) and (d), respectively. Here, three independent values, calculated based on the different $J=0,2,4$ coupling schemes~\cite{gosia2}, are presented for each level. 

The magnitude of the quadrupole asymmetry, $\left <\mathrm{cos}\; 3\delta \right >$, for levels in the ground-state and $\gamma$ bands is presented in Fig.~\ref{FIG4lab}. 
Oblate, triaxial, and prolate shapes correspond to $\left <\mathrm{cos}\; 3\delta \right >$ values of $-1$ ($\delta=60^\circ$), $0$ ($\delta=30^\circ$), and $1$ ($\delta=0^\circ$), respectively. The agreement between the four independent values of $\left <\mathrm{cos}\; 3\delta \right >$ indicates convergence of the present data set. Furthermore, the near constancy of this asymmetry parameter over the measured spin range for both bands confirms the presence of strong correlations between the $E2$ properties and, hence, the same deformation, as anticipated for collective behavior. Compared to the $\left < Q^2 \right>$ invariant, however, $\left <\mathrm{cos}\; 3\delta \right >$ appears to exhibit a small increase with spin, although a constant value is not ruled out within the quoted errors [Fig.~\ref{FIG4lab}(a)]. The average value of $\left <\mathrm{cos}\; 3\delta \right >\sim0.15$ for the ground-state band corresponds to a deformation $\expval{\delta}$ of $\sim 27^\circ$, in line with expectations for a well-defined triaxial shape. Within the quoted errors, the $\left <\mathrm{cos}\; 3\delta \right >$ behavior for the $\gamma$ band is the same [Fig.~\ref{FIG4lab}(b)], and the average value of 0.24 corresponds to a deformation parameter $\expval{\delta}$ of $\sim 25^\circ$. Hence, the quadrupole asymmetry, as determined from the $\left <\mathrm{cos}\; 3\delta \right >$ invariant, provides compelling evidence for triaxial deformation in both the ground-state and $\gamma$ bands, in agreement with the interpretation proposed in Ref.~\cite{Toh2013} based on the pattern reported for the energy staggering in the latter sequence. The present conclusions are also in line with calculations within the TPSM ($\gamma\approx 30^\circ$)~\cite{PhysRevC.89.014328} and SCCM ($\gamma\approx 25^\circ$)~\cite{Rodr_guez_2017} frameworks.

 \begin{figure}[t!]
\begin{center}
\hspace{-1.5cm}
\includegraphics[width=1.2\columnwidth]{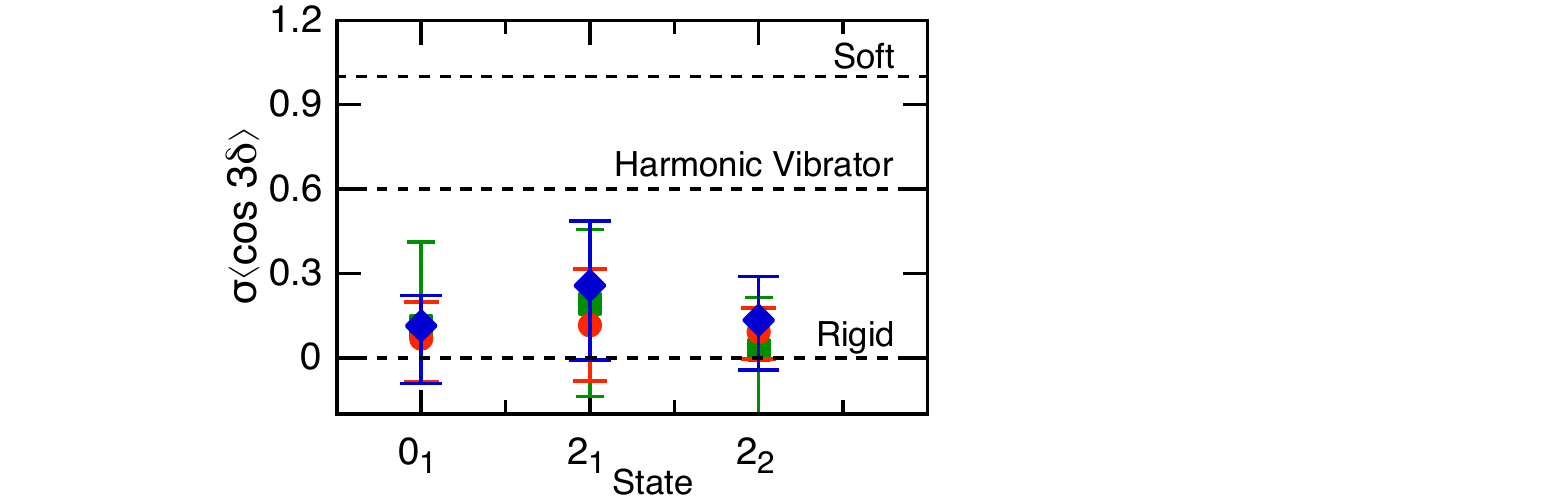}
\caption{(Color online) Statistical dispersion of the asymmetry for the $0^+_1$ and $2^+_1$ states in the ground band and the $2^+_2$ bandhead of the $\gamma$ band. Three independent measures of $\sigma\left <\mathrm{cos}\; 3\delta \right >$ are shown for each state. The same color convention as in FIG.~\ref{FIG3lab} has been used.}
\label{FIG5lab}
\end{center}
\end{figure}

As noted above, the nature of triaxial deformation can only be inferred from a higher-order invariant; e.g., the statistical fluctuation, or dispersion, $\sigma\left <\mathrm{cos}\; 3\delta \right >$ which determines the degree of rigidity - or softness - in the $\gamma$ degree of freedom. Figure~\ref{FIG5lab} presents the magnitude measured for this quantity from the present data for the three lowest states in $^{76}$Ge: the $0^+_1$ and $2^+_1$ levels in the ground-state band and the $2^+_2$ bandhead of the $\gamma$ band. It is worth noting that data on the $\sigma\left <\mathrm{cos}\; 3\delta \right >$ variance were also obtained for higher-spin states. However, these are not presented here as the incompleteness of the available data set likely increases with angular momentum, herewith resulting in an interpretation that cannot be proposed with the same degree of certainty. Nevertheless, the available data at these higher-spin values display a trend similar to that reported for the lowest-spin states in Fig.~\ref{FIG5lab}. By definition, rigid-triaxial deformation corresponds to values of $\sigma\left <\mathrm{cos}\; 3\delta \right >$ close to 0. In contrast, a harmonic vibrator is associated with values around 0.6, while a soft triaxial rotor is characterized by $\sigma\left <\mathrm{cos}\; 3\delta \right > \sim 1$~\cite{KK1,KK2,WU1996178}. The agreement between the three independent measures of $\sigma\left <\mathrm{cos}\; 3\delta \right >$  seen in Fig.~\ref{FIG5lab} for the three states of interest not only indicates convergence, but also strongly points to rigid triaxiality for $^{76}$Ge at and near its ground state. 

The present results indicate that the ground-state and $\gamma$ bands are characterized by the same $\beta$ and $\gamma$ deformation parameters as well as by the same degree of triaxial rigidity. Consequently, these observations directly impact the nuclear matrix elements relevant for neutrinoless double-beta decay: the various theoretical approaches will have to reproduce the parent $^{76}$Ge as a rigid triaxial rotor while also allowing for triaxiality in the $^{76}$Se daughter. In the latter case, the recent results of Ref.~\cite{PhysRevC.99.054313} indicate also a significant triaxiality with a dominant prolate component, but the degree of rigidity could not be determined. The potential might well be softer in this instance, as the odd-even staggering in the $\gamma$ band is opposite to that seen in $^{76}$Ge, but is in line with that reported for the other even Se and Kr isotopes of the region as well as for all other Ge isotopes with the exception of  $^{78}$Ge~\cite{PhysRevLett.120.212501}. 

In conclusion, a detailed study of the low-spin structure of the nucleus $^{76}$Ge has been undertaken following Coulomb excitation. An extensive and unique set of reduced $E2$ matrix elements was determined, enabling a model-independent analysis of the nature of triaxial deformation in $^{76}$Ge. Results of the rotational-invariant sum-rule analysis indicate, within experimental errors, that all levels observed in the ground-state and $\gamma$ bands are characterized by the same quadrupole ($\beta$) and asymmetry ($\gamma$) parameter values and, hence, are characterized by the same deformation. Most importantly, the results demonstrate that the low-spin structure of $^{76}$Ge is rigid triaxial, with a $\gamma$ value close to $30^\circ$. These conclusions are important for calculations aiming to determine the nuclear matrix elements relevant for $0\nu\beta\beta$ decay.

\begin{acknowledgments}
 The authors thank J. M. Allmond, J. Engel and B. Bally for valuable discussions. This work was funded by the U.S. Department of Energy, Office of Science, Office of Nuclear Physics, under Contracts No. DE-AC02-06CH11357 (ANL), DE-AC52-07NA27344 (LLNL), DE-AC02-05CH11231 (LBNL), and under Grant Numbers DE-FG02-97ER41041 (UNC), DE-FG02-97ER41033 (TUNL), DE-FG02-08ER41556 (MSU), DE-FG02-94ER40848 (UML), and DE-FG02-94ER4084 (Maryland) and by the National Science Foundation under grants Grants No. PHY-1565546, PHY-1811855 (MSU) and, PHY-1502092 (USNA). GRETINA was funded by the U.S. DOE, Office of Science, Office of Nuclear Physics under the ANL and LBNL contract numbers above. ADA acknowledges support provided by Office of Naval Research (ONR) through the Naval Academy Research Council (NARC). This research used resources of ANL’s ATLAS facility, which is a DOE Office of Science User Facility. 
\end{acknowledgments}

\bibliographystyle{apsrev4-1}
\bibliography{76Ge-Ayangeakaa-submit-V2}
\end{document}